\documentclass[useAMS,usenatbib]{mn2e}
\usepackage{amssymb}
\usepackage{graphicx}
\usepackage[fleqn]{amsmath}
\def\O{{\cal O}}
\def\b#1{\bmath#1}
\def\d{{\rm d}}

\def\i{{\rm i}}
\def\E{{\cal E}}

\def\p{\partial}

\long\def\crap#1{}
\def\=#1{{\langle#1\rangle}}

\title[Grommet]{{\sc Grommet}: an $N$-body code for high-resolution
  simulations of individual galaxies}
\author[S. J. Magorrian]{John Magorrian\thanks{E-mail:
    magog@thphys.ox.ac.uk}\\
  Rudolf Peierls Centre for Theoretical Physics, 1 Keble Road, Oxford
  OX1 3NP}
\begin{document}

\date{}

\volume{381}
\pagerange{1663--1671} \pubyear{2007}

\maketitle

\label{firstpage}

\begin{abstract}
  This paper presents a fast, economical particle-multiple-mesh
  $N$-body code optimized for large-$N$ modelling of collisionless
  dynamical processes, such as black-hole wandering or bar-halo
  interactions, occurring within isolated galaxies.  The code has been
  specially designed to conserve linear momentum.  Despite this, it
  also has variable softening and an efficient block-timestep scheme:
  the force between any pair of particles is calculated using the
  finest mesh that encloses them both (respecting Newton's third law)
  and is updated only on the longest timestep of the two (which
  conserves momentum).  For realistic galaxy models with
  $N\gtrsim10^6$, it is faster than the fastest comparable
  momentum-conserving tree code by a factor ranging from $\sim 2$
  (using single timesteps) to $\sim10$ (multiple timesteps in a
  concentrated galaxy).
\end{abstract}


\section{Introduction}

Newton's third law is a central pillar of physics.  Much of what we
know about the dynamical evolution of galaxies comes from $N$-body
simulation, but most $N$-body codes use approximations that break the
third law.
A well-known example of the consequences of breaking it is provided by
the sinking satellite problem \citep{HW89,W89,VW99}; the dynamical
friction felt by the satellite is grossly overestimated if one
``pins'' the centre of the host galaxy, ignoring the galaxy's $l=1$
dipole response.
This example is perhaps extreme, but there are many other situations
where one is interested in the detailed response of a galaxy to
asymmetric perturbations and would like to be able to model it without
having to worry about artifacts arising from violations of Newton's third law.
Examples include modelling bar-halo
interactions (see \citet{McMehnen} and references therein) and the
wandering of central supermassive black holes.

This paper describes the $N$-body code {\sc grommet} (GRavity On Multiple
Meshes Economically and Transparently), which has been designed
specifically to model the detailed dynamical evolution of individual
galaxies without using any approximations that violate Newton's third
law.  I assume that the galaxy is collisionless.  It is
completely described by a distribution function (DF) $f(\b x,\b v;t)$,
which gives the (mass) density of particles in phase space, along with
the potential~$\Phi(\b x;t)$ generated by this DF and any external
sources.  The evolution of $f$ is governed by the collisionless
Boltzmann equation (CBE),
\begin{equation}
  \label{eq:CBE}
  \frac{\p f}{\p t} + \b v\cdot\nabla f + \b a\cdot\frac{\p f}{\p\b v}=0,
\end{equation}
where the accelerations $\b a\equiv -\p\Phi/\p\b x$.  As \citet{HO92}
and \citet{LCB} emphasise, in a collisionless $N$-body code particles
are not to be thought of as representing stars or groups of stars.
Instead one is using the method of characteristics to
integrate~(\ref{eq:CBE}), estimating the accelerations $\b a(\b x)$ by
Monte Carlo sampling.  Of course, the shot noise in these estimates
means that in practice any simulation will never be perfectly
collisionless.  Therefore it is important to make $N$ as large as
possible in order to minimize the effects of this noise.  So, {\sc grommet}
has been designed to be both fast and economical on memory.

In section~\ref{sec:potsolve} below I describe the multiple-mesh
procedure used by {\sc grommet} to estimate accelerations.
Section~\ref{sec:move} shows how this leads naturally to a
momentum-conserving block-timestep integrator based on Duncan, Levison
\& Lee's (1998) potential-splitting scheme.  In
section~\ref{sec:tests} I present the results of some tests and also
compare {\sc grommet}'s performance against other codes'.
Section~\ref{sec:summary} sums up.  For completeness, I include in an
Appendix an explanation of James' (1977) method, which is used in
Section~\ref{sec:potsolve}.

\section[]{Potential solver}
\label{sec:potsolve}
The task of the potential solver in a collisionless $N$-body code is to
estimate the accelerations
\begin{equation}
  \label{eq:accels}
  \b a(\b x) = -\nabla\int {G\rho(\b x')\over |\b x-\b x'|}\,\d^3\b x',
\end{equation}
where one does not know the density distribution $\rho(\b x)$
explicitly, but instead only has a discrete sample of $N$ particles
with positions $\b x_i$ and masses $m_i$ drawn from it.

\subsection{Particle-mesh method}
At the heart of {\sc grommet}'s potential solver is the particle mesh (PM)
method \citep{Hock}.  It uses a cubical mesh, with vertices at
positions $\b x_{ijk}$, spaced a distance $h$ apart.  The procedure
for obtaining an initial estimate of the accelerations
(eq.~\ref{eq:accels}) felt by each particle follows.
\begin{enumerate}
\item Loop over all $N$ particles using cloud-in-cell interpolation
  to build up the discretized density distribution $\rho_{ijk}=\rho(\b
  x_{ijk})$;
\item Calculate the potential~$\Phi_{ijk}$
  corresponding to this~$\rho_{ijk}$ using James' (1977) method (see
  Appendix); 
\item Looping again over all $N$ particles, use a finite-difference
  approximation to estimate accelerations -$\p\Phi/\p\b x$ at the mesh
  points surrounding each particle, then interpolate the value of the
  acceleration at the particle's location using the same cloud-in-cell
  scheme employed in step~(i).
\end{enumerate}
Since steps (i) and~(iii) use the same interpolation scheme, this
procedure produces accelerations that obey Newton's third law subject
to one extra condition: the finite-difference scheme in step (iii)
cannot provide meaningful accelerations for particles that lie in the
outermost layer of mesh cells, which means that those particles should
be omitted in step~(i).  This seems an almost trivial point, but it is
important for the refinement scheme introduced below.  It turns out
that for the scheme below to work properly we have to peel off the
outer {\it two} layers of cells.  I typically use meshes with $64^3$
or $128^3$ cells, of which then only $60^3$ or $124^3$ are assignable
in step~(i).

Apart from respecting Newton's third law, the other attractive
features of the PM method are its efficiency and its linear scaling
with~$N$:  the time needed to carry out step (ii) is
independent of $N$, but for a typical mesh with $64^3$ cells the
overall time is dominated by the $\O(N)$ cost of carrying out the
assignment steps (i) and (iii) once $N\gtrsim5\times10^5$; similarly, the
memory needed to store mesh quantities and to carry out James' method
is negligible compared to that used for storing the particles'
masses, positions and accelerations.

The major disadvantage of the PM method is that it does not work well
for centrally concentrated mass distributions, since each particle has
an effective size of order the mesh spacing~$h$.  In other words, the
mesh spacing sets the effective softening length used in the
calculation of the forces.

\begin{figure}
\begin{center}\includegraphics[width=0.5\hsize]{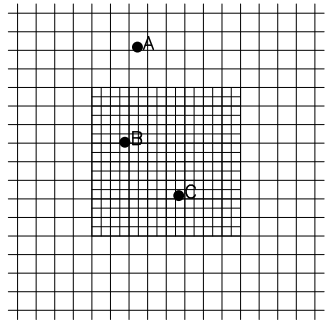}\end{center}
\caption{An example of the multiple mesh scheme used to calculate
  accelerations.  Particles A, B and~C all lie within the region
  covered by the outer, coarse mesh, but B and C also lie inside the
  fine, inner mesh.  An initial estimate of the forces on all three
  particles comes from using the PM method on the coarse mesh.  This
  is refined by isolating those particles within the inner mesh,
  recalculating their interparticle forces first using the fine mesh,
  then using the coarse, and adding the difference 
  to the initial coarse-mesh estimate.  Therefore, the force
  between A and each of B and C is obtained using the coarse mesh, but
  that between B and~C comes from the fine mesh.  In all cases
  Newton's third law is respected. \label{fig:mesh}}
\end{figure}
\subsection{Refinement scheme}

The natural remedy of this shortcoming is to introduce finer submeshes
in interesting, higher-density regions and to recalculate the
accelerations for particles inside each submesh.  But how best to
include the effect of the parent mesh's gravity field on the
accelerations calculated in each submesh and vice versa?  One
possibility is to solve Poisson's equation on the submesh subject to
boundary conditions interpolated from the parent mesh
\citep[e.g.,][]{Anninos,Jessop}.  This is a key element of the
widely-used family of multigrid methods \citep[e.g,.][]{ART,MLAPM},
and would be straightforward to apply in {\sc grommet} using the method of
equivalent charges (see Appendix).  However, all of these schemes
violate Newton's third law, as one can easily see by considering the
force between a particle inside a submesh and another one outside.

{\sc grommet} instead uses a simplified version of the scheme proposed by
\citet{Gelato} (see also figure~\ref{fig:mesh}).  The acceleration
felt by each particle is calculated using a series of nested
``boxes''.  We start with the outermost toplevel box, which
discretizes the simulation volume into, say, $n_x\times n_y\times
n_z=60^3$ assignable cells.
\crap{
The PM scheme
on this toplevel box provides an initial estimate of the particles'
accelerations.  
}
This box, like any other box, can contain zero, one or more subboxes.
Each subbox contains two meshes: a coarse one composed of an
$(n_x/2)\times (n_y/2)\times (n_z/2)$ subblock of the parent's cells,
and a fine one that covers the same subblock twice as finely in each
direction, with $n_x\times n_y\times n_z$ cells.  

For the most common situation in which each box contains no
more than one subbox, the acceleration at any position $\b x$ is given
by the sum over all boxes,
\begin{equation}
  \label{eq:grommetaccel}
  \b a(\b x) = \sum_j \b a_j(\b x),
\end{equation}
where the contribution from the $j^{\rm th}$ box,
\begin{equation}
\label{eq:al}
  \b a_j(\b x) = \b a_j^{+}(\b x) - \b a_j^{-}(\b x),
\end{equation}
is the difference between accelerations calculated using the PM method
on the box's fine (+) and coarse (-) meshes, simply ignoring any
particles that lie outside.  The outermost toplevel box
($j=0$) has no coarse mesh, so $\b a_0^{-}=0$.  In this scheme the
acceleration between any two particles is calculated using the box
with the finest mesh spacing that encloses them both and Newton's
third law is obeyed to machine precision.  This last feature comes at
a cost though: the acceleration~(\ref{eq:grommetaccel}) is
discontinuous at box boundaries, a point to which I return below.

Sometimes one might want to refine a region that cannot be
enclosed within a single subbox.  If one simply tiles the region
using, say, two abutting subboxes, the force between particles located
at either side of the boundary between them will be calculated using
the coarse parent mesh, which
is usually not desirable.  The solution is to let the subboxes overlap by
a few mesh cells and then correct eq.~(\ref{eq:grommetaccel}) for the
double counting of particles in the overlap region by introducing a
third subbox whose boundaries are given by the intersection of the two
overlapping subboxes and subtracting the accelerations~(\ref{eq:al})
obtained in this new subbox.  In contrast, \citet{Gelato} introduce a
buffer zone around each box and treat particles in the buffer zone
differently from the rest.  Their scheme violates Newton's third law.

I have deliberately omitted any automated scheme for deciding where
and when to introduce subboxes; these schemes inevitably break
time-reversibility, and, for the type of problem the code was designed
for, I expect that the user will already have a much better idea
of how best to place boxes.  

\section{Moving particles}
\label{sec:move}

The characteristic equation of the CBE is
\begin{equation}
  \label{eq:charac}
  \frac{\d t}{1} = \frac{\d\b x}{\b v} = \frac{\d\b v}{\b a},
\end{equation}
where the accelerations $\b a(\b x,t)$ depend on the DF~$f$ through
equ.~(\ref{eq:accels}).  The most straightforward and widely used way
of following the characteristics is by using a leapfrog integrator.
The (fixed-timestep) leapfrog produces an approximate solution
to~(\ref{eq:charac}) that respects many of its important symmetries;
it is symplectic\footnote{This assumes that the accelerations are
  smooth, which is not the case for many collisionless
  $N$-body codes, including {\sc grommet}.},
reversible in time and, when the accelerations are
obtained using a potential solver that respects Newton's third law,
it conserves linear momentum.

An unattractive feature of the leapfrog is that it uses the same fixed
timestep for all particles.  Consider a deeply plunging radial orbit
in a model galaxy with a central density cusp or black hole.
Integrating this orbit accurately near pericentre requires a very
small timestep, which, in the standard leapfrog scheme, means that all
other particles have to be integrated using the same small timestep,
even those on loosely bound circular orbits.  This can be
prohibitively expensive, since it involves calculating the full set of
accelerations $\b a(\b x,t)$ for all particles at every timestep.

{\sc grommet} uses a block-timestep scheme to improve efficiency.  Each of
the boxes of section~\ref{sec:potsolve} above has an associated
timestep, which can be chosen to be either equal to that of its parent
box or a factor of two shorter.
Broadly speaking, a particle's
position and velocity are updated using the shortest timestep of any
of the boxes enclosing it, but the force between any pair of particles
is updated only on the timestep of the longest particle, thus
conserving linear momentum.  The rest of this section makes this
somewhat vague description more precise.

\begin{figure*}
\begin{center}
\includegraphics[width=0.8\hsize]{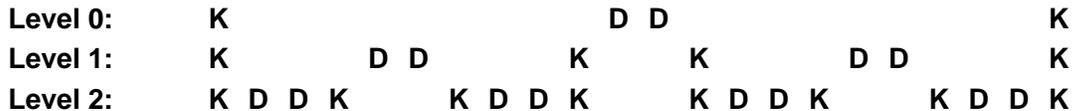}
\end{center}
\caption{The sequence of steps for motion in the
  Hamiltonian~(\ref{eq:hamfrogs}) with two levels of timestep
  refinement.
  For any given timestep level~$l$, the $K$ operation ``kicks''
  particles inside any boxes having that timestep level, applying to each an
  impulse $\frac12\tau_l\cdot(-\p V_{(l)}/\p\b x)$, where the timestep
  $\tau_l=2^{-l}\tau_0$.
  These impulses change the particles' velocities, but not their
  positions.  They conserve the particles' total linear momentum.
  The $D$ operation ``drifts'' all particles
  for a time $\frac12\tau_l$, changing their positions but not their velocities.
  \label{fig:multikddk}}
\end{figure*}
\subsection{The standard leapfrog integrator}
Recall that a leapfrog integrator with a single, fixed timestep $\tau$
corresponds to motion in a time-dependent Hamiltonian
\citep[e.g.,][]{Hotbert}
\begin{equation}
\label{eq:hamfrog}
  H = T + \sum_{k=-\infty}^\infty \delta_\epsilon\left(k-\frac{t}{\tau}\right) V(\b
      x_1,\ldots,\b x_N),
\end{equation}
where $T\equiv\frac12\sum_i m_iv_i^2$ is the kinetic energy of all the
particles and
$\delta_\epsilon(x)\equiv\frac12(\delta(x-\epsilon)+\delta(x+\epsilon))$
with $0<\epsilon\ll1$.  The periodic comb of delta functions turns on
the potential energy $V(\b x_1,\ldots,\b x_N)$ only at times
$t=(k\pm\epsilon)\tau$ for integer~$k$.  Integrating the
resulting equations of motion from time $t=k\tau$ to $t=(k+1)\tau$
yields
\begin{align}
  \b v_i(k+\textstyle\frac12) &= \b v_i(k)+
             {\textstyle\frac12}\tau\b a_i(k),\\
  \b x_i(k+1) & = \b x_i(k) + \tau \b v_i(k+\textstyle\frac12),\\
  \b v_i(k+1) & = \b v_i(k+\textstyle{\frac12})+\textstyle{\frac12}\tau\b a_i(k+1),
\end{align}
where the accelerations $\b a_i(k)\equiv-{\p V}/m_i{\p\b x_i}$
evaluated at time~$t=k\tau$.  This is just the sequence of steps for
the kick-drift-kick form of the leapfrog: the potential is turned on
briefly just after $t=k\tau$ resulting in a ``kick'' (denoted $K$) to the
particles' velocities; the particles then ``drift'' ($D$) along at
their new velocities until the potential turns on again just before
$t=(k+1)\tau$, at which point they receive another kick.  The
drift-kick-drift form of the leapfrog can be obtained by adding
$\frac12$ to the argument of the delta functions or, alternatively, by
integrating the equations of motion from $(k-\frac12)\tau$ to
$(k+\frac12)\tau$ instead.

Another way of looking at each of these versions of the leapfrog is to
consider them as compositions of the two time-asymmetric first-order
symplectic integrators (each applied left to right),
$K(\tau/2)D(\tau/2)$ and $D(\tau/2)K(\tau/2)$, whose first-order error
terms cancel \citep[e.g.,][]{SahaTremaine}.  In the following I write
the leapfrogs as the sequence of operations $KDDK$ and $DKKD$,
dropping the $(\tau/2)$ arguments.

\subsection{A block-timestep leapfrog}
\label{sec:multimove}
In {\sc grommet} the accelerations $\b a(\b x)$ are given by a
sum~(\ref{eq:grommetaccel}) of contributions~(\ref{eq:al}) from boxes
with different spatial refinement levels.  The outermost box is
associated with a timestep $\tau_0$ and timestep level $l=0$.
Each subbox has a timestep $\tau_l=2^{-l}\tau_0$ with timestep
level $l$ either equal to that of its parent or larger by one.
Let us add together all the
contributions~(\ref{eq:al}) to $\b a(\b x)$ from boxes having timestep
level~$l$ and write the result as $\b a_{(l)}(\b x)$.  Let
$V_{(l)}(\b x)$ be the corresponding contribution to the potential
energy.
Instead of turning on the full potential $V=\sum_l V_{(l)}$ at every
timestep, consider the alternative Hamiltonian
\begin{equation}
\label{eq:hamfrogs}
  H = T + \sum_{l=0}^{l_{\rm max}}
\sum_{k=-\infty}^\infty \delta_\epsilon\left(k-\frac{t}{2^{-l}\tau_0}\right) V_{(l)}(\b
      x_1,\ldots,\b x_N),
\end{equation}
where $l_{\rm max}$ is the maximum timestep refinement level and each
$V_{(l)}$ is turned on only at times $t=2^{-l}k\tau_0$.  This is a
variant of the potential splitting used by \citet{DLL98} to model
close encounters in planetary systems.

Integrating the equations of motion for this new Hamiltonian results
in a nested sequence of $KDDK$ leapfrog steps, as shown in
figure~\ref{fig:multikddk}.  The sequence can be produced using the
following simple recursive algorithm:
\vbox{
\obeylines
\tt
Step($l$, $\tau$):
\quad if $l>l_{\rm max}$:
\qquad   Drift($\tau/2$)
\quad else:
\qquad Kick($l$,$\tau/2$)
\qquad  Step($l+1$,$\tau/2$)
\qquad  Step($l+1$,$\tau/2$)
\qquad Kick($l$,$\tau/2$)
}
This algorithm is called initially with $l=0$ and $\tau=\tau_0$.  Each
{\tt Kick($l$,$\tau/2$)} operation applies an impulse
$-\frac12\tau\nabla V_{(l)}$ to all particles, which changes the
particles' velocities but not their positions.  The {\tt Drift}
operation moves the particles once the complete set of impulses has
been applied.

This algorithm requires a factor $\sim l_{\rm max}$ fewer kick
operations (and therefore fewer expensive force evaluations) than a
simple leapfrog with a single timestep $2^{-l_{\rm max}}\tau_0$.
It is obvious that it conserves linear momentum and is reversible in
time.  Unlike the integrator in \citet{DLL98}, however, it is {\it
  not} symplectic; the discontinuities in the accelerations at box
boundaries mean that the Poincar\'e integral invariants are not
conserved.

\section{Tests and comparisons}
\label{sec:tests}

I have carried out a number of simple tests with small numbers of
particles ($1<N\lesssim20$) to confirm that my implementation of the
ideas above really does respect Newton's third law and conserve linear
momentum.  These small-$N$ tests serve only as minimal sanity checks;
as stressed by \cite{MLAPM}, truly interesting tests of a
collisionless code come not from testing how faithfully it reproduces
the solution to the two-body problem, but rather from its
ability to model collisionless systems accurately using large numbers
of particles.

In this section I use some simple collisionless galaxy models to test
{\sc grommet}'s potential solver and integrator, comparing results obtained
from {\sc grommet} against those obtained from two other codes.  Both of the
other codes are available as part of the NEMO package.  The first is
the fast tree code described in \citet{Dehnen02}.  It obtains
accelerations from a Cartesian multipole expansion.  This respects
Newton's third law and a standard leapfrog integrator built around
this potential solver then naturally conserves linear momentum.  (A
multiple-timestep version is also available, but it does not conserve
momentum.)  The second code \citep{HO92} uses the so-called
``self-consistent field'' (SCF) method, which represents the density
and potential using a truncated basis function expansion.  It shows no
respect for Newton's third law, but, like {\sc grommet}, is optimized for
modelling single galaxies.

\begin{figure}
\begin{center}\includegraphics[width=0.8\hsize]{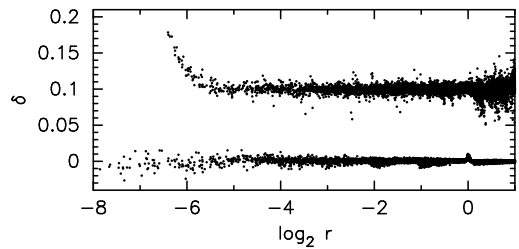}\end{center}
\caption{Fractional errors in the accelerations 
  at randomly selected positions within and around an
  $N=10^7$-particle realization of a
  truncated power-law sphere.  The lower set of points plot results
  calculated using the potential solver of section~\ref{sec:potsolve}
   using 8 levels of refinement
  of a $60^3$ mesh with $x_{\rm max}=2$.
  The upper set (offset by 0.1 vertically) are for
  results obtained using a tree code with
  fixed softening length 
  $\epsilon=10^{-2}$.
  \label{fig:fracaccels}}
\end{figure}

\begin{figure}
\begin{center}\includegraphics[width=0.8\hsize]{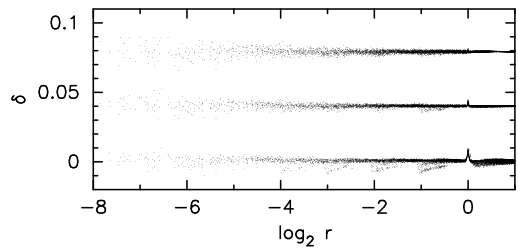}\end{center}
\caption{Fractional errors in the accelerations inside a
  $10^8$-particle realization of a power-law sphere, a factor of 10
  more particles than in figure~\ref{fig:fracaccels}.  The lower set
  of points plot results obtained using the potential solver of
  section~\ref{sec:potsolve} with the same set of nested boxes and
  $60^3$ mesh employed for figure~\ref{fig:fracaccels}.
  The middle and upper set show the effects of using finer meshes with
  $124^3$ (middle) and $252^3$ (upper) cells, offset by 0.04 and 0.08 respectively.
\label{fig:fracaccelszoom}}
\end{figure}

\subsection{Static tests}

Real galaxies have steep central density cusps
\citep[e.g.,][]{LauerNuk}, so an obvious test of the potential solver
is to check the accelerations it returns for an $N$-body realization
of a truncated power-law sphere with density profile
\begin{equation}
\rho(r)\propto
\begin{cases}
  r^{-\alpha}, & \hbox{if $r<r_{\rm max}$},\cr
  0,     & \hbox{otherwise.}
\end{cases}
\end{equation}
I have generated a realization with $r_{\rm max}=1$, $\alpha=2$ having
$10^7$ equal-mass particles and used eq.~(\ref{eq:grommetaccel}) to
calculate accelerations at randomly selected positions inside and
around the sphere.  For this I use a toplevel box enclosing the
region $|\b x|<2$ together with eight levels of refinement, the
boundary of the $i^{\rm th}$ subbox being given by $|\b x|=2^{1-i}$.
Figure~\ref{fig:fracaccels} plots the fractional difference between
the results of this procedure against the exact, analytical expression
for the acceleration.  For radii $2^{-5}<r<1$ the RMS fractional
error is only 0.0023, rising to 0.007 for $2^{-8}<r<2^{-5}$, within
which there are relatively few particles.  The source of this good and
desirable behaviour is the decrease in the effective softening length
as one moves to smaller length scales.

For comparison, the upper set of points in figure~\ref{fig:fracaccels}
plot the errors in the accelerations of the same $10^7$-particle
sphere calculated at the same positions using the tree code {{\sc falcon}}
with softening kernel $P_2$ and fixed softening length
$\epsilon=10^{-2}$.  The RMS fractional error in the resulting
accelerations for radii $2^{-5}<r<1$ is 0.011, over four times larger than
{\sc grommet}'s, while for $r<2^{-5}$, the calculated accelerations become
systematically too low.  {{\sc falcon}} takes about 2.5 times longer than
{\sc grommet} to produce these results and needs more than three times the
memory.

Perhaps the most worrying feature of the nested box scheme of
section~\ref{sec:potsolve} is that the
accelerations~(\ref{eq:grommetaccel}) are discontinuous at box
boundaries.  One can see some hints of this discontinuity in
figure~\ref{fig:fracaccels} at $\log_2 r = -1$, $-2$, $-3$, but it is
even clearer in figure~\ref{fig:fracaccelszoom} which plots the
fractional errors in a $10^8$-particle realization.
Even if one were to run a simulation with such large $N$, the
discontinuity itself is unlikely to be important because the
integration scheme in section~\ref{sec:move} does not depend
explicitly on the derivatives of the accelerations (but the
discontinuity does mean that the integrator is not symplectic, as
noted earlier).  More important is the fact that if the discontinuity
is noticeable it means that the bias in the estimates of the
accelerations has become significant.  The natural solution is then to
move to a finer mesh (e.g., $124^3$ cells instead of $60^3$,
figure~\ref{fig:fracaccelszoom}).

\subsection{Dynamical tests}
\label{sec:hernqtest}
For the dynamical tests I use a spherical isotropic \citet{Hernquist}
model with density profile
\begin{equation}
  \label{eq:hernq}
  \rho(r) = \frac{Ma}{2\pi r(a+r)^3}.
\end{equation}
This idealized model is in equilibrium.  Then by Jeans' theorem
\citep{BT} its DF~$f_0(\b x,\b v)$ can depend on $(\b x,\b v)$ only
through the integrals of motion, which are the energy $\E$ and angular
momentum $\b J$ per unit mass.  Since the model is isotropic the DF
cannot depend on the latter and so $f=f_0(\E)$.

A straightforward procedure for generating initial conditions
(hereafter ICs) corresponding to this model would be to draw $N$
particles directly from $f_0(\E)$, assigning each a mass $M/N$.
Integrating~(\ref{eq:hernq}), the fraction of particles inside
radius~$r$ would then be $r^2/(a+r)^2$, showing that there would be
relatively few particles with radii $r\ll a$, deep inside the
interesting $r^{-1}$ central density cusp.  To improve resolution near
the centre, I instead generate initial conditions using a multi-mass
scheme, drawing particles from an anisotropic
sampling DF \citep{LCB} with {\it number} density
\begin{equation}
  \label{eq:fs}
f_s(\E,J^2) = h(\E,J^2)f_0(\E),
\end{equation}
where \citep{Sigurdsson} 
\begin{equation}
 h(\E,J^2) \equiv A\times
  \begin{cases}
    \left(\frac{r_{\rm peri}}a\right)^{-\lambda} & \hbox{if $r_{\rm peri}<a$},\cr
    1 & \hbox{otherwise},
  \end{cases}
\end{equation}
$r_{\rm peri}(\E,J^2)$ is the particle's pericentre radius
and the constant $A$ is chosen to normalize~$f_s$.
When the parameter $\lambda=0$, the sampling DF $f_s$ is identical to
$f_0(\E)$.  Increasing $\lambda$ improves the sampling of the cusp by
increasing the number density of particles having pericentres $r_{\rm
  peri}<a$.  To balance this increase in number density each particle
is assigned a mass $Mf_0/Nf_s=M/Nh(\E,J^2)$ so that the phase-space
mass density is still given by the desired~$f_0(\E)$

For the tests below I adopt units $G=M=a=1$ and draw $2\times10^6$
particles with radii in the range $10^{-3}<r<10^2$ from the sampling
DF~(\ref{eq:fs}) with $\lambda=1$.  Poisson noise in the resulting
distribution of particles makes it slightly asymmetric, which has two
unwanted consequences \citep[see also][]{McMehnen}.  First, the centre
of mass of the system moves with a constant velocity of order
$\sim10^{-3}(GM/a)^{1/2}$ because the total linear momentum of the
particles is small, but non-zero.  Second, the asymmetry quickly
destroys the inner part of the $r^{-1}$ density cusp, even when viewed
a frame co-moving with the centre of mass.  To remove both of these
effects, I extend my ICs to include the mirror distribution obtained
by reflecting each of the $2\times10^6$ particles with $(\b x,\b
v)\to(-\b x,-\b v)$.  The full ICs then have $N=4\times10^6$
particles.

\begin{figure}
\begin{center}
\includegraphics[width=0.8\hsize]{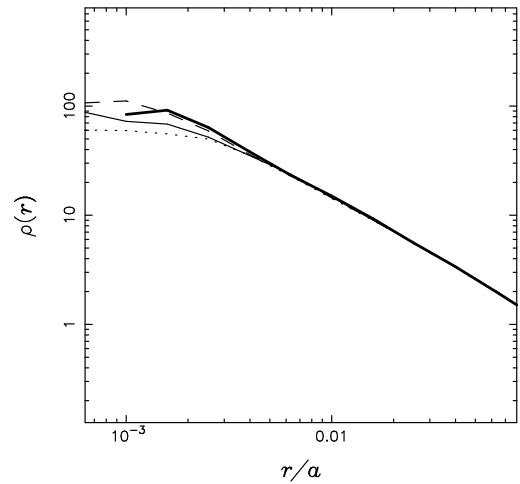}
\end{center}
\caption{Inner density profile of the same realization
  of a Hernquist model after it has been evolved for 10 time units
  using a simple leapfrog integrator with
  accelerations obtained using different potential solvers: {\sc grommet} (light solid curve), {\tt
    {\sc falcon}} (dotted curve) and the SCF method (dashed).  The heavy solid curve
  plots the density profile of the the initial conditions.
  \label{fig:hernqrho}}
\end{figure}

\begin{figure*}
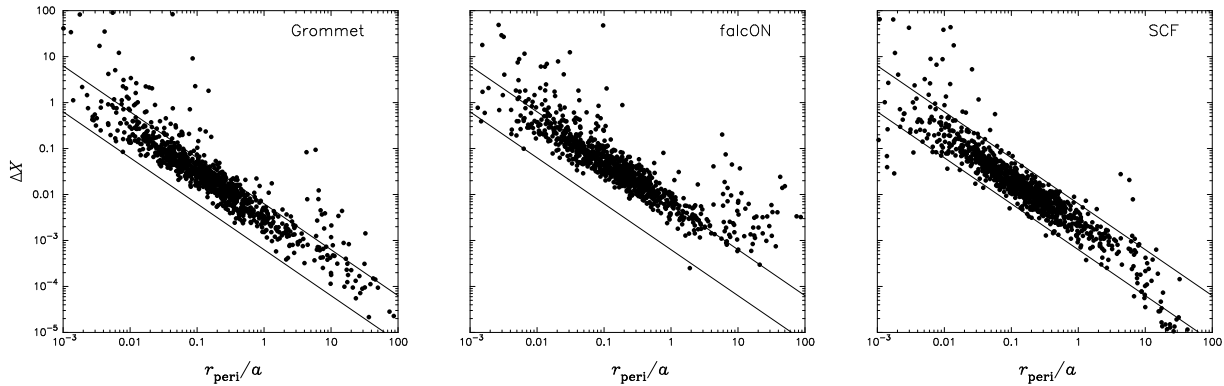

\begin{center}
\includegraphics[width=0.3\hsize]{diffXgrommet}
\includegraphics[width=0.3\hsize]{diffXfalcon}
\includegraphics[width=0.3\hsize]{diffXscf}
\end{center}
\caption{RMS fractional change in the angular momenta
of particles in the models
  of figure~\ref{fig:hernqrho}, measured from $t=0$ to $t=10$
  and plotted as a function of the particles' pericentre radii.  The same random
  selection of particles is used to generate each panel. 
\label{fig:hernqdiff}}
\end{figure*}

\subsubsection{Evolution of an (almost) equilibrium model}

Of course, one does not expect an $N$-body model evolved from these
ICs to be in perfect equilibrium; the ICs omit particles outside the
range $10^{-3}<r<10^2$ and are constructed assuming the exact
potential corresponding to the density distribution~(\ref{eq:hernq})
instead of the softened potential used in the $N$-body code.
Nevertheless, it is interesting to compare the evolution of the
$N$-body model obtained from {\sc grommet} with those obtained from the
other two codes.

Figure~\ref{fig:hernqrho} shows the density profile of the models
after 10 time units (or $\sim66$ circular orbit periods at $r=0.01$).
All three models use the same simple leapfrog integrator with timestep
$2\times10^{-3}$; only the source of the accelerations is different.
For {\sc grommet} I use boxes with boundaries at $|\b x|=100\times2^{-i}$
for $i=0,\ldots,12$.  Each box has $60^3$ assignable cells, the cell
length varying from 3.33 in the toplevel box down to
$0.8\times10^{-3}$ in the innermost box.  {\sc falcon}'s results are
obtained using kernel $P_2$ with softening length $\epsilon=10^{-3}$,
while the SCF expansion uses the \citet{HO92} basis function expansion
truncated at $n_{\rm max}=6$ radial and $l_{\rm max}=4$ angular terms.

The results in figure~\ref{fig:hernqrho} are unsurprising.  The
density at the very centre of the {\sc grommet} and {\sc falcon} models falls
slowly because because the ICs omit particles with radii $r<10^{-3}$
and do not take into account the softening in these codes.  In
contrast, the density profile of the SCF model does not change
significantly because its basis function expansion is incapable of
producing anything that deviates strongly from a Hernquist model on
small spatial scales.

Much more is happening at the level of individual orbits, however.
All of these models begin with spherical symmetry and remain
spherical, apart from the effects of Poisson noise.  Therefore the
amount of diffusion in the angular momentum~$J$ of their particles'
orbits serves as a convenient measure of how far each code is from
being perfectly collisionless.  
Figure~\ref{fig:hernqdiff} shows that the particles in all three
models suffer from significant amounts of diffusion.  The SCF model
shows the least diffusion, but it is only marginally better
than {\sc grommet}; although the SCF potential remains close to the exact
Hernquist potential, the flickering of the expansion coefficients with
time makes the orbits diffuse just like in any other code.  The
diffusion is worst in the {\sc falcon} model, particularly for orbits having
pericentres much larger than its fixed softening length
$\epsilon=10^{-3}$.  All of these results are based on the variation
in orbits' angular momentum in models integrated from $t=0$ to
$t=10$, but I find similar results for models integrated from, say, $t=10$ to
$t=50$ when scaled to account for the longer timescale over which the
diffusion occurs.

\begin{table}
  \centering
  \begin{tabular}{rcl}
\bf Code & \bf Time & \bf Comment\\
{\sc falcon} & 2.1 & single timestep\\
SCF & 1.3 & single timestep, $(n_{\rm max},l_{\rm max})=(6,4)$ \\
{\sc grommet} & 1.0 & single timestep\\
{\sc grommet}  & 0.3 & four levels of timestep refinement\\
{\sc grommet} & 0.16 & seven levels of timestep refinement\\
\end{tabular}
  \caption{Comparison of time required for different codes to integrate the multi-mass
    Hernquist model of section~\ref{sec:hernqtest}, relative to the 
    single-timestep implementation of {\sc grommet}.
    Neither the {\sc grommet} nor the SCF models take advantage of the reflection symmetry of
    this simple problem.}
  \label{tab:timings}
\end{table}

The results presented so far have been obtained using an integrator
with a single small timestep, but the dynamical time inside the cusp
of a Hernquist model varies with radius $r$ approximately as
$r^{1/2}$.  As, e.g., \citet{Zemp} have argued, it is natural to
advance particles using a timestep proportional to the local dynamical
time.  We can come close to the optimal $\tau\propto r^{1/2}$ scaling by
using the block-timestep integrator of section~\ref{sec:multimove}
above and halving the timestep on every {second} subbox.
To test the practicality of this scheme, I have run a model with
timesteps $\tau=32\times10^{-3}$ for particles with $|\b
x|>100\times2^{-6}\simeq1.5$, shrinking by a factor of two at the
boundaries $|\b x|=100\times2^{-i}$ of boxes $i=6$, 8, 10 and~12.  In
the innermost ($i=12$) box the timestep is $2\times10^{-3}$, the same
used for the single-timestep run above.  This multiple-timestep model
yields results almost indistinguishable from the single-timestep {\sc grommet}
model plotted in figures \ref{fig:hernqrho} and~\ref{fig:hernqdiff},
but is three times faster (see table~\ref{tab:timings}).  If it were
appropriate to halve the timestep at {\it all} box boundaries
$i=6,\ldots,12$ (see below for an example) then the block-timestep
scheme would yield a sixfold increase in speed over the
single-timestep integrator.

\begin{figure}
\begin{center}
\includegraphics[width=0.8\hsize]{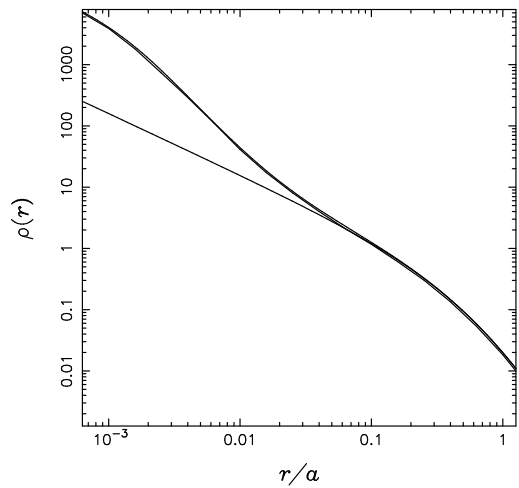}
\end{center}
\caption{The results of adiabatically adding a Plummer sphere
 potential to an initially isotropic Hernquist model.
  The Plummer sphere has radius $2\times10^{-3}a$ and a final mass
  $2\times10^{-3}\,M_{\rm gal}$.  
The results obtained using {\sc grommet}'s multiple-timestep scheme are almost
 identical to those calculated from Young's (1980) method.
  \label{fig:hernqadia}}
\end{figure}

\subsubsection{Response to an adiabatically grown blob}

For a slightly more interesting test, I model the growth of a black
hole at the centre of a galaxy by slowly adding a Plummer sphere
potential
\begin{equation}
  \label{eq:plummer}
  \Phi_{\rm b}(\b x; t) = -\frac{GM_{\rm b}(t)}{\sqrt{r^2+b^2}}
\end{equation}
to a multi-mass Hernquist model.  The scale radius of the Plummer
sphere $b=2\times10^{-3}$ and its mass grows with time as \citep{Sigurdsson}
\begin{equation}
  \label{eq:massplum}
  M_{\rm b}(t) = M_{\rm f}\times
  \begin{cases}
   \left[ 3\left(\frac{t}{t_{\rm g}}\right)^2 - 
        2\left(\frac{t}{t_{\rm g}}\right)^3 \right] & \hbox{if
        $t<t_{\rm g}$}\\
       1  & \hbox{otherwise},
  \end{cases}
\end{equation}
its final mass $M_{\rm f} = 2\times10^{-3}$ being reached in a time $t_{\rm g}=5$.  

A safe, formal way of including the effects of this external potential
in {\sc grommet} is to add an extra term
\begin{equation}
\sum_{k=-\infty}^\infty \delta_\epsilon\left(k-\frac{t}{2^{-l_{\rm
        max}}\tau_0}\right)\sum_{i=1}^N m_i\Phi_{\rm b}(\b x_i;t)
\end{equation}
to the Hamiltonian~(\ref{eq:hamfrogs}).  Integrating the resulting
equations of motion then leads to the modifications needed in the
block-timestep algorithm (section~\ref{sec:multimove}).  In this case
the necessary modifications are obvious, but for more realistic
situations (e.g., if the mass of the external source did not change in
time and if the location of its centre were not pinned to $\b x=0$)
then it is helpful to start from~(\ref{eq:hamfrogs}) to ensure that
the perturbation is turned on at the appropriate times and momentum
conserved.

As above, I use a nested series of boxes with boundaries at $|\b
x|=100\times2^{-i}$ with $i=0,\ldots,12$, each box covered by a $60^3$
mesh.  Boxes 0 to~5 share a common timestep~$\tau_0=5\times10^{-3}$.
This is refined in every subsequent box, so that the timestep
associated with box~$i\ge6$ is $2^{5-i}\tau_0$ and the innermost box
($|\b x|<0.024$) has timestep $\sim4\times10^{-5}$.

My initial conditions consist of $10^6$ particles drawn from the
sampling DF~(\ref{eq:fs}) above.  The artificially imposed potential at
$\b x=0$ means that this simulation only makes sense if the particles'
centre of mass is also at $\b x=0$.  As an alternative to symmetrizing
the ICs as before, I instead modify step~(i) of the PM method
(section~\ref{sec:potsolve}) to reflect the particle distribution
through each of the planes $\b x=0$, $\b y=0$, $\b z=0$ when assigning
mass to meshes.  This increases the effective $N$ used for the
potential by a factor of 8 at little cost.  The density profile of the
final model is plotted in figure~\ref{fig:hernqadia}.  It agrees well
with the predictions obtained using Young's (1980) method.

\section{Summary}
\label{sec:summary}

I have described {\sc grommet}, a fast, economical particle-multiple-mesh
$N$-body code designed specifically for modelling the dynamical
evolution of individual galaxies.  In other words, it is designed to
tackle almost exactly the same type of problem to which the SCF method
\citep{HO92} is applied.  Indeed, {\sc grommet} can -- loosely -- be thought
of as a variant of the SCF method using a Cartesian basis function
expansion with millions of expansion coefficients (the density at each
mesh vertex in each of the nested boxes).  Any application of the SCF
method requires that one make a careful choice of the basis functions
used to represent the density and potential.  Similarly, in {\sc grommet}
one has to choose, by hand, the set of nested boxes to use.

For a realistic model galaxy with $N\gtrsim10^6$, the single-timestep
incarnation of {\sc grommet} is comparable in speed to an SCF code using a
low-order basis expansion and shows comparable amounts of relaxation.
For most applications, however, {\sc grommet} will be much faster: its
nested-box potential solver admits an efficient natural block-timestep
integrator (section~\ref{sec:multimove}), leading to an approximate
three- to six-fold increase in speed for realistic galaxy models; the
SCF method typically requires a fairly high-order expansion to produce
(reasonably) unbiased results \citep[e.g.,][]{KHB}, which makes it
much slower in practice.  But perhaps the main advantage of {\sc grommet}
over SCF methods based on spherical harmonic expansions is that it
respects Newton's third law and is therefore suitable for use in
studying $l=1$ perturbations without fear of artefacts due to
centring.

To my knowledge, the tree code {\sc falcon} \citep{Dehnen02} is the only
other code that can model realistically inhomogeneous galaxies without
breaking the third law.  For $N\gtrsim10^6$ {\sc grommet}'s potential solver
is more than twice as fast as {\sc falcon}'s and much less memory hungry.
This efficiency comes at a cost though, since {\sc grommet}'s nested-box
scheme is optimized for modelling perturbations of single galaxies.
It would be interesting to see whether the potential-splitting scheme
used here (section~\ref{sec:multimove}; \citet{DLL98}) works as well
for {\sc falcon}, or indeed any other code that respects the third law, as
it does for {\sc grommet}.

\section*{Acknowledgments}
I thank James Binney, Walter Dehnen and Ben Moore for
helpful discussions, and the Royal Society for financial support.

\appendix

\section[]{James' method}

\label{app:James}
\citet{James} describes an economical method for calculating the solution
to Poisson's equation,
\begin{equation}
  \label{eq:poisson}
  \nabla^2\Phi = - q,
\end{equation}
discretized on a regular mesh and with a spatially bounded source
distribution~$q(\b x)$, so that $\p\Phi/\p\b r\to0$ as $r\to\infty$.
It is easiest to explain his method for the electrostatic case in
which $q$ is electric charge density and $\Phi$ is electrostatic
potential.  The method then consists of the following steps:
\begin{enumerate}
\item enclose the charge distribution $q$ inside an earthed metal box
  and calculate the potential $\phi(\b x)$ inside the box subject
  to the boundary condition $\phi=0$ on the box surface;
\item use Gauss' law to find the charge distribution $Q$ induced on
  the surface of the box;
\item calculate the potential $\psi(\b x)$ due to this
  surface charge distribution $Q$.
\end{enumerate}
The solution to~(\ref{eq:poisson}) for the isolated charge
distribution $q$ is given by $\Phi = \phi - \psi$.
Since this procedure is at the heart of {\sc grommet}'s potential solver, I
explain it in some detail below.

\subsection{Preliminaries}

Throughout the following, I assume that the distribution $q(\b x)$ has
been discretized onto a cubical mesh with vertices at positions $\b
x_{ijk}=(i,j,k)$, $0\le i,j,k\le n$, spaced unit distance apart.  Our
goal is to calculate the discretized potential $\Phi_{ijk}$
corresponding to this $q_{ijk}$.

A straightforward way of doing this is to use the Fourier convolution
theorem.  Consider first the situation in which the charge
distribution $q_j$ is one-dimensional with $0\le j<2n$; the reason for
extending the mesh from $n+1$ to $2n$ vertices will become apparent
shortly.  The discretized potential is given by the convolution
\begin{equation}
  \label{eq:onedconv}
  \Phi_j = \sum_{k=0}^{2n-1}G_k q_{k-j},
\end{equation}
where $G_k$ is the contribution to $\Phi_k$ made by a unit-charge
particle located at $x_0$, and we take $q_k=0$ for $k<0$
or $k\ge 2n$ since we have an isolated charge distribution.

This last condition on $q$ is awkward.  Suppose instead that both that
$G_k$ and $q_k$ were $2n$-periodic, with $q_{-k}=q_{2n-k}$, and let us
impose the sensible condition that $G_k$ is even with $G_k=G_{2n-k}$
and that $G_0=0$.  Then $\Phi_j$ could be obtained economically using
Fourier methods.  The Fourier transform of $q_i$ is given by
\begin{equation}
  \label{eq:stdft}
q^\alpha \equiv\sum_{j=0}^{2n-1}q_j\exp\left[\i\pi j\alpha\over n\right],
\end{equation}
where $\i\equiv\sqrt{-1}$, and similarly for $G^\alpha$.  Using the
discrete convolution theorem, equation~(\ref{eq:onedconv}) becomes
simply
\begin{equation}
  \label{eq:onedconvft}
  \Phi^\alpha = G^\alpha q^\alpha.
\end{equation}
Applying the inverse transform, the potential is given by
\begin{equation}
  \label{eq:inverseft}
  \Phi_j = {1\over 2n}\sum_{\alpha=0}^{2n-1} \Phi^\alpha
\exp\left[-{\i\pi j\alpha\over n}\right].
\end{equation}
The periodicity needed for application of the discrete
convolution theorem is a nuisance, but if we allow $q_i$ to be
non-zero only for $0\le i\le n$, then the $\Phi_0\ldots\Phi_n$
obtained from equ.~(\ref{eq:onedconv}) are unaffected by it.
Therefore, we can use this Fourier method to obtain the potential
$\Phi_i$ corresponding to an isolated source distribution $q_i$ ($0\le
i\le n$) provided we extend the mesh by a further $n-1$ points with
$q_{n+1}=\cdots=q_{2n-1}=0$.  Thanks to the existence of fast methods
for evaluating the transforms (\ref{eq:stdft})
and~(\ref{eq:inverseft}), the Fourier method requires only $\O(n\log
n)$ operations to calculate the full set of $\Phi_i$, instead of the
$\O(n^2)$ involved in a direct evaluation of equ.~(\ref{eq:onedconv}).
The savings are much more dramatic for the three-dimensional case, for
which the direct sum takes $\O(n^6)$ operations, compared to only
$\O((n\log n)^3)$ for the Fourier method.

James' method makes use of an alternative view of this ``doubling up''
procedure.  The Fourier transform~(\ref{eq:stdft}) can be written as
\begin{equation}
\label{eq:sincossplit}
  q^\alpha = 2\big[q^\alpha(C)+\i q^\alpha(S)\big],
\end{equation}
where the cosine and sine transforms
\begin{align}
  q^\alpha(C) & \equiv \sum_{j=0}^n c_jq_j \cos{\pi j\alpha\over n},\\
  q^\alpha(S) & \equiv \sum_{j=1}^{n-1} q_j\sin{\pi j\alpha\over n},
\end{align}
come from the even and odd parts of $q_j$,
\begin{equation}
  q^\pm_j \equiv {1\over2}\Big(q_j\pm q_{2n-j}\Big),
\end{equation}
respectively.
The coefficients $c_1=\cdots =c_{n-1}=1$, but $c_0=c_n={1\over2}$ to
account for the fact that $q_0$ and $q_n$ are counted only half as
often as the other $q_i$.  Conversely, having both $q^\alpha(C)$ and
$q^\alpha(S)$ we can reconstruct the original $q_j$ by
substituting~(\ref{eq:sincossplit}) into the expression for the
inverse transform~(\ref{eq:inverseft}) to obtain
\begin{equation}
  \label{eq:qsplitsc}
  q_j = q_j(C) + q_j(S),
\end{equation}
where
\begin{align}
  q_j(C) & \equiv {2\over n}
         \sum_{\alpha=0}^n c_\alpha
          q^\alpha(C) \cos{\pi j\alpha\over n},\\
  q_j(S) & \equiv {2\over n}
         \sum_{\alpha=1}^{n-1} q^\alpha(S)\sin{\pi j\alpha\over n}
  \label{eq:qsplitscbla}
\end{align}
are the inverse cosine and sine transforms of $q^\alpha(C)$ and
$q^\alpha(S)$ respectively.  Thus, apart from a factor of $(2/n)$, the
cosine and sine transforms are their own inverses.

Now suppose that only $q_0\ldots q_n$ are allowed to be non-zero.
Then $q^+_i=q^-_i=q_i$, except for the unused $q^-_0=q^-_n=0$.
Replacing $q$ by $\Phi$ in
equs. (\ref{eq:qsplitsc}-\ref{eq:qsplitscbla}) and taking
$\Phi^\alpha$ from~(\ref{eq:onedconvft}), the
potential can be written as
\begin{equation}
  \Phi_j = \Phi_j(C) + \Phi_j(S),
\end{equation}
where
\begin{align}
  \Phi_j(C) & \equiv {2\over n}
         \sum_{\alpha=0}^n c_\alpha
          G^\alpha q^\alpha(C) \cos{\pi j\alpha\over n},\\
  \Phi_j(S) & \equiv {2\over n}
         \sum_{\alpha=1}^{n-1} G^\alpha q^\alpha(S)\sin{\pi j\alpha\over n},
\end{align}
and $G^\alpha=2G^\alpha(C)$ with no contribution from the sine
transform of the even function $G_i$.

The generalization to three dimensions is straightforward.  The
Fourier transform in each direction splits into a sum of cosine and
sine terms, yielding a total of eight terms:
\begin{align}
\label{eq:phispliteight}
  \Phi_{ijk} & = \Phi_{ijk}(CCC) + 
\Phi_{ijk}(CCS) + 
\Phi_{ijk}(CSC) \nonumber \\ \nonumber
& \qquad + \Phi_{ijk}(CSS) + 
 \Phi_{ijk}(SCC) + 
\Phi_{ijk}(SCS) \\
& \qquad+ \Phi_{ijk}(SSC) + 
\Phi_{ijk}(SSS),
\end{align}
where, for example,
\begin{align}
\Phi_{ijk}(CSS) &
 \equiv
{8\over n^3}\sum_{i=0}^{n} \sum_{j=1}^{n-1}\sum_{k=1}^{n-1}
c_i \Phi^{\alpha\beta\gamma}(CSS) \times \\
&\qquad  \cos{\pi i\alpha\over n} \sin{\pi j\beta\over n}
      \sin{\pi k\gamma\over n},\nonumber
\end{align}
with
$\Phi^{\alpha\beta\gamma}(CSS)=G^{\alpha\beta\gamma}q^{\alpha\beta\gamma}(CSS)$
and 
\begin{equation}
  \label{eq:qcss}
  q^{\alpha\beta\gamma}(CSS)  \equiv
\sum_{i=0}^{n} \sum_{j=1}^{n-1}\sum_{k=1}^{n-1} c_iq_{ijk}
 \cos{\pi i\alpha\over n} \sin{\pi j\beta\over n}
      \sin{\pi k\gamma\over  n}.
\end{equation}
Notice that this decomposition into sine and cosine transforms results
in two transforms for each of eight $n^3$ meshes.  It requires less
memory than the equivalent single $(2n)^3$ zero-padded mesh used in
the ``doubling-up'' procedure, but for general $q_{ijk}$ and $G_{ijk}$
it offers no improvement in speed.  It simplifies enormously, however,
for the special case in which $G_{ijk}$ is the Green's function of the
discretized Laplacian appearing in equ.~(\ref{eq:poisson}).  James'
method exploits these simplifications, particularly in dealing with
the hollow induced surface charge distribution~$Q$ (see
section~\ref{sec:hollowpot} below).

\subsection{The potential of a charge distribution inside an earthed
  box}
\label{sec:earthedbox}
With this background in hand, let us turn to the details of James' method.
Poisson's equation~(\ref{eq:poisson}) can be written
\begin{equation}
  \label{eq:discretepoisson}
({\Delta\Phi})_{ijk}=- q_{ijk}
\end{equation}
where the first-order approximation to the Laplacian operator
\begin{align}
  \label{eq:L1}
(\Delta\Phi)_{ijk}\equiv
\Phi_{i+1,j,k}+\Phi_{i-1,j,k}
+\Phi_{i,j+1,k}+\Phi_{i,j-1,k}+\Big.\nonumber \\
\quad +\Phi_{i,j,k+1}+\Phi_{i,j,k-1}
-6\Phi_{ijk}.
\end{align}
The potential $\phi_{ijk}$ of the earthed box is given by the solution
of this equation subject to the boundary condition
$\phi_{0jk}=\phi_{njk}=\phi_{i0k}=\phi_{ink}=\phi_{ij0}=\phi_{ijn}=0$.
Applying the triple sine transform, we have that
\begin{equation}
  \label{eq:phisss}
  \phi^{\alpha\beta\gamma}(SSS) = q^{\alpha\beta\gamma}(SSS)/C^{\alpha\beta\gamma}
\end{equation}
where
\begin{equation}
  C^{\alpha\beta\gamma} = 2\left(1-\cos{\pi\alpha\over n}\right)
 + 2\left(1-\cos{\pi\beta\over n}\right)
 + 2\left(1-\cos{\pi\gamma\over n}\right).
\end{equation}
Although we could immediately apply the inverse
transform~(\ref{eq:phispliteight}) to obtain $\phi_{ijk}$ explicitly,
it turns out that this is unnecessary and it is more efficient to use
the method of equivalent charges (see below) to modify
$\phi^{\alpha\beta\gamma}(SSS)$ to include the effects of the
potential $\psi$ corresponding to the induced surface charge
distribution~$Q$, saving everything for a single inverse transform at
the very end of the calculation.

\subsection{The charge distribution induced on the faces of the box}
\label{sec:facecharges}
The charge distribution induced on the $i=0$ face of the box is
given by
\begin{equation}
  \label{eq:gauss}
  Q_{0jk} = -(\Delta{\phi})_{0jk} = -\phi_{1jk},
\end{equation}
the last equality following from the fact that $\phi$ is zero both on
the box surface ($i=0$) and outside the box ($i=-1$).  Similarly, the
charge distribution induced on the opposite $i=n$ face is $Q_{njk} =
\phi_{n-1,jk}$.  Notice that $Q$ vanishes along the edges of the box,
and so is completely specified by its double sine transform on each of
the six faces.  In terms of $\phi^{\alpha\beta\gamma}(SSS)$ these can
be written
\begin{align}
Q_0^{\cdot\beta\gamma}(SS) & = {2\over n}\sum_{\alpha=1}^{n-1}
    \phi^{\alpha\beta\gamma}(SSS)\sin{\pi\alpha\over n},\\
Q_{n}^{\cdot\beta\gamma}(SS) & = {2\over n}\sum_{\alpha=1}^{n-1}
    (-1)^\alpha \phi^{\alpha\beta\gamma}(SSS)\sin{\pi\alpha\over n},
\end{align}
and similarly for the other four faces.  We can invert each of these
to obtain $Q_{0jk}$ etc and, from these, any of the other three
transforms $Q_0^{\cdot\beta\gamma}(SC)$, $Q_0^{\cdot\beta\gamma}(CC)$,
$Q_0^{\cdot\beta\gamma}(CS)$.

\subsection{The potential of the induced charge distribution}
\label{sec:hollowpot}
Equation (\ref{eq:phispliteight}) provides a way of obtaining
the potential $\psi_{ijk}$ corresponding to this induced charge
distribution $Q$.  The result is a sum of eight terms, all of
which can be treated in the same way.  For example, consider the term
$\psi_{ijk}(CSS)$.  Its Fourier transform
$\psi^{\alpha\beta\gamma}(CSS)=G^{\alpha\beta\gamma}Q^{\alpha\beta\gamma}(CSS)$,
where, using~(\ref{eq:qcss}) and the hollowness of~$Q$,
\begin{equation}
  Q^{\alpha\beta\gamma}(CSS) = Q_0^{\cdot\beta\gamma}(SS) +
  (-1)^\alpha Q_n^{\cdot\beta\gamma}(SS).
\end{equation}
The other terms can be written in a similar way, although the $SSS$
term vanishes.  The $G^{\alpha\beta\gamma}$ used here should be the
triple cosine transform of the Green's function for the first-order
Laplacian~(\ref{eq:L1}).  This need be calculated just once (e.g.,
using the doubling-up procedure), the necessary elements being stored
for subsequent use.

It would be possible to use eq.~(\ref{eq:phispliteight}) to obtain
$\psi_{ijk}$ directly, but it turns out (see below) that this labour
is unnecessary and that it suffices to use~(\ref{eq:phispliteight}) to
obtain only the face potentials $\psi_{0jk}$, $\psi_{njk}$ etc.
Nevertheless, adding up all the contributions to each of these turns
out to be the main computational burden of James' method.

\subsection{The method of equivalent charges}

Instead of synthesizing $\psi_{ijk}$ explicitly, let us introduce
another potential $\psi^{(E)}_{ijk}$ which is zero on the box faces
but everywhere else is equal to $\psi_{ijk}$.  Because it vanishes at
the box boundaries, this new potential is completely specified by its
triple sine transform.  The ``equivalent charge'' distribution
$E_{ijk}$ that generates it can be found using Poisson's
equation
\begin{equation}
  \nabla^2\left[\psi-\psi^{(E)}\right]=-\left[Q-E\right],
\end{equation}
where $Q$ is non-zero only on the faces of the box.  For the
first-order discretized Laplacian (equ.~\ref{eq:L1})
$E$ is confined to the planes $i=1$, $j=1$, $k=1$, $i=n-1$, $j=n-1$
and $k=n-1$, with, for example,
$E_1^{\cdot\beta\gamma}(SS)=\psi_0^{\cdot\beta\gamma}(SS)$.

The triple sine transform of the full potential is then
\begin{equation}
\label{eq:sssfullpot}
  \Phi^{\alpha\beta\gamma}(SSS) =
  \phi^{\alpha\beta\gamma}(SSS)-E^{\alpha\beta\gamma}(SSS)/C^{\alpha\beta\gamma}
\end{equation}
the second term giving the contribution of $\psi^{(E)}$.  Applying the
inverse triple-sine transform to~(\ref{eq:sssfullpot}) gives
$\Phi_{ijk}$ for $1\le i,j,k<n$.  Finally, the missing face potentials
can be inserted using the results obtained in
section~\ref{sec:hollowpot}.

\subsection{Performance}

My implementation of this procedure uses the fast sine and cosine
transforms written by Takuya Ooura.\footnote{{\tt
    http://kurims.kyoto-u.ac.jp/\rlap{\hbox{\~\
      }}\phantom{\~}ooura/fft.html}}
The triple-sine transforms involved in going from $q_{ijk}$ to
$\phi^{\alpha\beta\gamma}(SSS)$ (eq.~\ref{eq:phisss}) and from
$\Phi^{\alpha\beta\gamma}(SSS)$ to $\Phi_{ijk}$
(eq.~\ref{eq:sssfullpot}) then take only $\sim30\%$ of the total time
needed to go from $q_{ijk}$ to~$\Phi_{ijk}$, with the evaluation of
the various transforms of the face charge distributions
(Sec.~\ref{sec:facecharges}) accounting for a further 10\%.  The
remaining 60\% of the time is spent simply accumulating the various
contributions to the face potentials (sec.~\ref{sec:hollowpot}).
Nevertheless, for typical $65^3$ or $129^3$ meshes I find that my
implementation of James' method is about 60 to 70\% faster than the
usual doubling-up procedure.

I have focused here on describing James' method using the first-order
approximation of the Laplacian (\ref{eq:L1}).
\cite{James} shows how it is possible to apply the same ideas to
higher-order approximations, albeit at the expense of much more
involved book-keeping.  I find that the resulting minor changes in the
Green's function $G_{ijk}$ have no detectable effect for the realistic
large-$N$ situations described in section~\ref{sec:tests}.

\end{document}